\documentclass[sigconf]{acmart} %,authordraft, ,anonymous

\usepackage{microtype}
\usepackage{listings}
\usepackage{tabulary, lipsum}
\usepackage{svg}
\usepackage{todonotes}
\usepackage{mathtools}
\usepackage{enumitem}
\usepackage{xspace}
\RequirePackage[nolist,nohyperlinks]{acronym}

\begin{acronym}
  \acro{RBAC}[RBAC]{Role Based Access Control}
  \acro{DDSA}[DDSA]{Data-Driven Software Architecture}
  \acro{DSL}[DSL]{domain-specific language}
  \acro{FODA}[FODA]{Feature-Oriented Domain Analysis}
  \acro{ADL}[ADL]{Architectural Description Language}
  \acro{DFD}[DFD]{Data Flow Diagram}
  \acroplural{DFD}[DFDs]{Data Flow Diagrams}
  \acro{ADD}[ADD]{Architectural Design Decision}
  \acro{LDA}[LDA]{Linear Discriminant Analysis}
  \acro{PCA}[PCA]{Principal Component Analysis}
  \acro{EFA}[EFA]{Exploratory Factor Analysis}
  \acro{LR}[LR]{Linear Regression}
  \acro{RF}[RF]{Random Forests}
  \acro{LGR}[LGR]{Logistic Regression}
  \acro{DFA}[DFA]{Data Flow Analysis}
  \acro{PCM}[PCM]{Palladio Component Model}
  \acro{ABUNAI}[ABUNAI]{Architecture-Based and Uncertainty-Aware Confidentiality Analysis}
  \acro{arcovia}[ARCoViA]{Automated Repair of Confidentiality Violations in Software Architectures}
  \acro{colja}[COLJA]{Compliance‑driven Joint Interdisciplinary Software Architecture Modeling and Analysis}
  \acro{TFG}[TFG]{Transposed Flow Graph}
  \acroplural{TFG}[TFGs]{Transposed Flow Graphs}
  \acro{GQM}[GQM]{Goal-Question-Metric}
  \acro{GDPR}[GDPR]{General Data Protection Regulation}
  \acro{ZTA}[ZTA]{Zero Trust Architecture}
  \acro{llm}[LLM]{Large Language Model}
\end{acronym}

\newcommand{\xdecaf}{{\footnotesize x}\textsc{decaf}\xspace}

\AtBeginDocument{%
  }

\setcopyright{acmlicensed}
\copyrightyear{2026}
\acmYear{2026}
\acmDOI{XXXXXXX.XXXXXXX}
\acmConference[ASE '26]{41st IEEE/ACM International Conference on Automated Software Engineering}{ October 12–-16,
  2026}{Munich, Germany}
\acmISBN{978-1-4503-XXXX-X/2018/06}

\begin{document}

\title{xDECAF: An Extensible Data Flow Diagram Analysis Framework for Information Security}

\author{Benjamin Arp}
\email{benjamin.arp@kit.edu}
\affiliation{%
    \institution{Karlsruhe Institute of Technology}
    \country{Germany}}
\author{Felix Schwickerath}
\email{felix.schwickerath@student.kit.edu}
\affiliation{%
    \institution{Karlsruhe Institute of Technology}
    \country{Germany}}
\author{Alexander Vogt}
\email{alexander.vogt@student.kit.edu}
\affiliation{%
    \institution{Karlsruhe Institute of Technology}
    \country{Germany}}
\author{Tom H{\"u}ller}
\email{tom.hueller@student.kit.edu}
\affiliation{%
    \institution{Karlsruhe Institute of Technology}
    \country{Germany}}
\author{Nils Niehues}
\email{nils.niehues@kit.edu}
\affiliation{%
    \institution{Karlsruhe Institute of Technology}
    \country{Germany}}
\author{Nicolas Boltz}
\email{nicolas.boltz@kit.edu}
\affiliation{%
    \institution{Karlsruhe Institute of Technology}
    \country{Germany}}

\renewcommand{\shortauthors}{Arp et al.}

\begin{abstract}
\xdecaf is an extensible tool for architecture-based data flow analysis with a focus on information security.
It combines an extended data flow diagram metamodel of labeled flows and nodes, a domain-specific constraint language with different flow operations, and a browser-based editor backed by an analysis engine.
In this paper, we present the \xdecaf tool library and a curated catalog of over 20 example models with documented constraints and expected violations, intended as a reusable dataset for the community. 
The tool has already been adopted by several research lines, providing concrete evidence of its utility. 
The tool, dataset, and a hosted online editor are publicly available.

\noindent\textbf{Screencast:} \url{https://youtu.be/L-PGdWoPtlw}\\
\noindent\textbf{Code \& Dataset:} \url{https://github.com/DataFlowAnalysis} \textbullet{} \newline\url{https://doi.org/10.5281/zenodo.20083877}
\end{abstract}

%    <concept>
%        <concept_id>10002978.10003022.10003023</concept_id>
%        <concept_desc>Security and privacy~Software security engineering</concept_desc>
%        <concept_significance>500</concept_significance>
%        </concept>
%  </ccs2012>

% \ccsdesc[500]{Security and privacy~Software security engineering}

\keywords{Data Flow Diagram, Data Flow Analysis, Propagation, Constraint Language, Example Catalogue, Tooling, Web Editor}


\maketitle
%%When reading through prior submission the tool demo looks more like a user documentation style then a highly academic submission. Things like user group, interfaces, features and usage are often emphasised

\section{Introduction}
\label{sec:intro}

Modern software systems increasingly shape critical parts of everyday life and are rarely confined to a single technical or organizational setting. They commonly span organizational boundaries, heterogeneous technologies, and multiple regulatory contexts, which makes it necessary to assess them systematically against security and compliance requirements.
To reason about such properties at design time, \acp{DFD} are a common notation for representing the structure of relevant parts of a system \cite{shostack2008experiences,owasp-threat-dragon}.

In this tool paper, we present \xdecaf, a \ac{DFD}-based extensible framework for security modeling and automated analysis, with a focus on information security.
\xdecaf is intended as a foundation for researchers who want to address data-flow-based analysis problems.
They can use its existing capabilities as building blocks or adapt individual aspects of the framework and tooling to support more specialized or advanced research goals.

To support this intended use, the core of \xdecaf was designed around a general data-flow analysis paradigm. It supports user-defined semantics by allowing label annotations, propagation logic, and data flow constraints to be freely specified. This enables analyses beyond singular security concerns and across different application domains.
Reflecting this generality, the tooling surrounding its core is designed for reuse and extension. The analysis provides explicit interfaces using a builder pattern, allowing it to be integrated as an analysis component (e.g., as an oracle) or combined with domain-specific pre-/postprocessing. Core pipeline stages can also be exchanged or reimplemented, e.g., import formats and their interpretation, or data flow identification. \xdecaf further provides an extensible online editor with frontend and backend components to support outward-facing, end-user-oriented research, such as improved threat-modeling workflows.
% Um Platz zu sparen könnte dies hier in eine Art Outline umformuliert werden, bzw. mit der Outline verbunden werden
% To support this intended use, \xdecaf was designed with customizability and extensibility in mind. Its core analysis supports arbitrary semantic properties via label annotations and freely definable data propagation logic, making it applicable beyond individual security properties or compliance domains. In addition, the analysis is designed as a reusable library with explicit interfaces, enabling domain-specific use in different research prototypes and application scenarios, e.g., as an oracle for information security properties or in usage with specific preprocessing components. \todo{Was ist unser unique selling point (klarer herausstellen)}
%and \todo{or as an oracle for information security properties} postprocessing components.
%\xdecaf further provides an online editor with frontend and backend components that can be reused and extended for outward-facing, end-user-oriented research, such as improving threat-modeling support.

The artifacts presented in this tool and dataset paper are:
\begin{itemize}[leftmargin=1.5em, itemsep=0em, topsep=3pt]
	\item the \xdecaf tool library, with interfaces for reuse and extension,
	\item the online editor with frontend and backend, and
	\item a catalog of example models, provided as a reusable dataset.
\end{itemize}

The utility and validation of these artifacts is supported by research that builds on \xdecaf.
The underlying analysis foundations of \xdecaf, including its scalability, have been described and validated in prior work \cite{boltz2024extensible}. Subsequent research has used \xdecaf to address additional security-related properties \cite{boltz2024modeling}, to integrate further architectural modeling languages and \ac{DFD} datasets \cite{seifermann2022detecting,microSecEnD23,niehues2024integrating,huller2024towards}, and to support analysis composition \cite{reiche2025detecting}, uncertainty modeling and analysis \cite{hahner2023impact}, as well as automated mitigation of data-flow violations \cite{niehues2025architecture,niehues2026efficient}. Beyond information security, \xdecaf has also been applied to legal compliance scenarios \cite{boltz2025knowledge,boltz2026continuous}.
\xdecaf and all artifacts described in this paper are open-source and linked together with the artifacts listed above. In addition, we provide a dataset that captures the current release of all described artifacts \cite{dataset}. 

\section{\xdecaf Core Concepts}
\label{sec:concepts}
% xDECAf is made up of several... Extended DFD syntax, propagation logic, contraint language + analyse)
%\todo{kurzer einleitender Satz basierend auf dem Kommentar oben.}
The core of \xdecaf is made up of an extended DFD syntax, propagation logic, constraint language, and analysis engine.
These concepts have been presented and evaluated in detail in previous work~\cite{boltz2024extensible}.

\begin{figure}[htbp]
  \centering
  \vspace{-10pt}
  \setlength{\abovecaptionskip}{5pt}
  \setlength{\belowcaptionskip}{0pt}
  \includegraphics[width=\columnwidth]{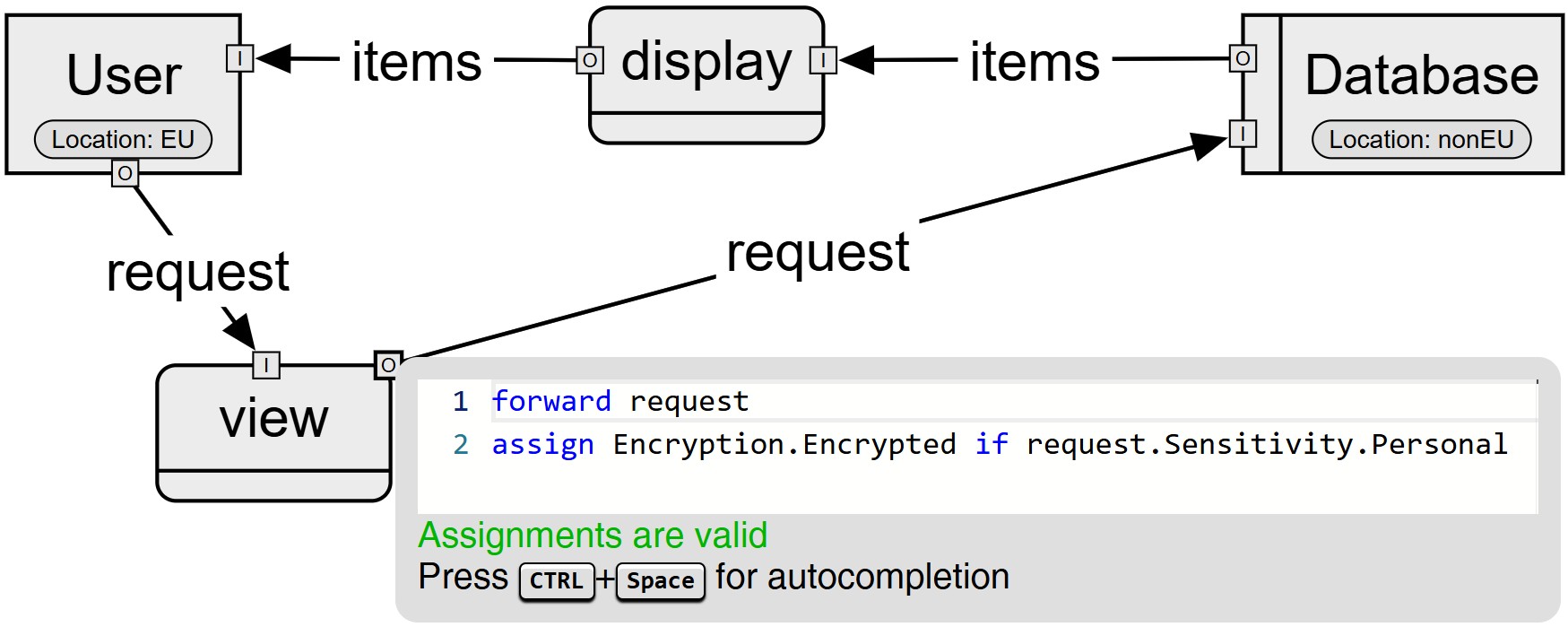}
  \caption{Extended DFD Example in xDECAF online editor.}
  \vspace{-10pt}
  \label{fig:extended-dfd-example}
\end{figure}

%The data flow diagrams (DFDs) of \xdecaf are based on the notation of \citet{demarco1979} \todo{Demarco raus und hier einfach wie unten im kommentar} but extend it with \emph{Labels}, \emph{Pins}, and \emph{Assignments} to facilitate the core label propagation approach.
The \acp{DFD} of \xdecaf extend the basic notation of nodes and flows with \emph{Labels}, \emph{Pins}, and \emph{Assignments} to facilitate the core label propagation approach.
\emph{Labels} represent discrete values and are used to annotate specific properties/metainformation to either nodes or to data propagated through the system. \emph{Labels} assigned to nodes may represent node-specific properties, e.g., deployment information or security levels. \emph{Data Labels} are propagated through the system and represent metainformation about the data, e.g., sensitivity, granularity, or required access levels.
Pins are assigned directly to a node and act as the input and output interfaces of said node. Therefore, as shown in \autoref{fig:extended-dfd-example}, flows between nodes start and terminate in \emph{Pins}. 
Multiple flows into the same \emph{Input Pin} represent multiple independent paths data can take to reach that node, while multiple flows into different \emph{Output Pin} of the same node represent parallel flows. %\todo{hier korrekt Satz beenden}
\emph{Assignments} are part of \emph{Output Pins} and describe how \emph{Labels} are propagated along their connected flows. They allow the formulation of conditional logical statements that reference input pins and labels to define whether, or under which conditions, incoming data \emph{Labels} are modified.
After modeling, the propagation logic of \xdecaf propagates the labels through the flows of the DFD according to the assignment specifications. 
The label propagation process identifies all possible routes that label information can take through the system. As a result, \xdecaf identifies all independent data flows that connect $n$ sources to a sink. If these data flows contain cycles, heuristics are applied to resolve them \cite{arp2024analyzing}.

To analyze the propagated DFDs, \xdecaf provides a \ac{DSL}\footnote{\url{https://dataflowanalysis.org/wiki/dsl/} (11.05.2026)} for formulating and evaluating data flow constraints. Constraints are evaluated independently for each individual data flow.
The constraint language expresses constraints by describing the origin and destination of flows and their relationship in the following structure: {\small \texttt{<Source> <flowVerb> <Destination>}} and an additional optional section for additional constraints denoted with {\small\texttt{where <Conditional>}}.
Both the {\small \texttt{<Source>}} and {\small \texttt{<Destination>}} sections allow users to define selectors that match elements as sources or destinations. Possible selectors include matching labels of a node or data, as well as the node type, node name, or data name.
Additionally, selectors can be combined using logical operators, such as \texttt{\&} or \texttt{|}.
Each {\small \texttt{<flowVerb>}} is interpreted as a quantifier over the set $X$ of flows $x$ between a matched source $x_{src}$ and a matched destination $x_{dst}$: the predicate $\varphi(x) \coloneqq\varphi(x_{src},x_{dst})$ holds when the selectors are matched. 
We support four \emph{flow verbs}, corresponding to the two quantifiers and their negations of first-order logic: $\textit{flows} \equiv \exists x ~\varphi(x)$, $\textit{alwaysFlows}  \equiv \forall x ~\varphi(x)$, $\textit{neverFlows}  \equiv \lnot\exists x ~\varphi(x)$,\\ $\textit{notAlwaysFlows} \equiv \lnot\forall x ~\varphi(x)$.
\begin{comment}The \texttt{<flowVerb>} describes the way the constraint is evaluated and can be categorized into the following:
\begin{table}[h]
  \footnotesize
  \centering
  \label{tab:flow-verbs}
  \begin{tabular}{@{}ll@{}}
    \toprule
    \textbf{Verb} & \textbf{Semantics over data flow paths $\pi$} \\
    \midrule
    \texttt{flows}          & $\exists\pi.\; \mathit{src}$ reaches $\mathit{dst}$ on $\pi$ \\
    \texttt{alwaysFlows}    & $\forall\pi.\; \mathit{src}$ reaches $\mathit{dst}$ on $\pi$ \\
    \texttt{neverFlows}     & $\forall\pi.\; \mathit{src}$ does not reach $\mathit{dst}$ \\
    \texttt{notAlwaysFlows} & $\exists\pi.\; \mathit{src}$ does not reach $\mathit{dst}$ \\
    \bottomrule
  \end{tabular}
  \caption{...}
\end{table}
\todo{needs description}
\end{comment}
Lastly, in the {\small \texttt{<Conditional>}} section, relationships between the source and destination sections can be enforced. 
This is achieved by introducing \emph{Variables} in the source and destination sections that contain all possible propagated label values.
These sets of label values can be used in selectors that implement set operations, like intersection and size operations.
A role-based access control constraint, with \emph{Conditional} and \emph{Variables}, is shown in \autoref{lst:constraint:role}. It defines that data should only be accessed by a certain granted role, should only flow to nodes with the correct permissions, as denoted by assigned roles.

\begin{lstlisting}[
    float,
    mathescape=false,
    belowcaptionskip=-15pt,
    caption={xDECAF constraint for role-based access control.},
    label={lst:constraint:role}
]
 data
    with GrantedRoles.$grantedRoles 
 never flows to
 node
    with assignedRoles.$assignedRoles
 where empty intersection($grantedRoles,$assignedRoles)
\end{lstlisting}

\xdecaf also supports the analysis of \ac{PCM} instances. The \ac{PCM} \cite{palladioBook} is an \ac{ADL} for component-based software systems. An extension of the \ac{PCM} allows the definition of confidentiality-related properties of components, interfaces, and resource descriptions. A transformation extracts \acp{DFD} from the control-flow modeled in \ac{PCM} instances and uses the confidentiality-related properties as labels in the analysis of \xdecaf \cite{seifermann2022detecting, huller2024towards}.

%% Tom ~ 0.3

%% Felix ~  0.3
\section{Example Model Catalogue}
%% Frame as a reusable dataset
In addition to the analysis outlined in the previous sections, we provide a reusable dataset \cite{dataset} that includes models, documentation, data flow constraints, and expected violations. 
The catalog includes models of different levels of complexity and several domains, which are either derived from other literature in the information security context \cite{seifermann2019,tuma_flaws_2019,tuma2021,palladioBook,van2025privacy,smartspeaker}, from available documentation \cite{hahner2023impact,ccc2024VWCariad,heinrich2024cocar}, industry collaboration \cite{corallo2026everestdataset} and/or are inspired by recent exploits \cite{tuma_flaws_2019,ccc2024VWCariad}.
The provided models illustrate security pitfalls and support researchers as a well-documented foundation for evaluation.
Overall, the catalog contains 26 models that range from 7 to 923 nodes and from 4 to 72 individual labels.
The complexity of models depends on several factors, mainly the amount of independent data flows and included cycles. For smaller models (e.g., > 20 nodes), \xdecaf takes > 1 second to complete an analysis run, and does not exceed 1 minute for our most complex models from the catalog.
This collection complements the previously integrated {microSecEnD} dataset \cite{microSecEnD23, niehues2024integrating}, containing security-annotated DFDs derived from the source code of open-source microservice applications. 
A list of all current models is also found on the \xdecaf website\footnote{\url{https://dataflowanalysis.org/examples/models/} (11.05.2026)}.

%% all(Tom,Felix)? ~ 0.5
\section{Online Editor} %% 0.5

%% Frontend + Backend
%% List key features: modeling, constraints, results.
%% Add one screenshot ?? (no)

To allow easy creation, visualization, and analysis of DFDs without requiring code or a local installation, we provide a publicly available web-based editor\footnote{\url{https://editor.dataflowanalysis.org/} (11.05.2026)}.
The editor is split into a frontend for visualization and a backend that runs the analysis and converts from different inputs.
The frontend is built on the Eclipse Sprotty\footnote{\url{https://sprotty.org/} (11.05.2026)} framework, which manages diagram rendering and maintains the graphical model of the DFD on an interactive SVG-based canvas. Surrounding this canvas are several floating UI panels, including a palette for creating \acp{DFD}, a label editor for managing labels, and a constraint editor for defining DSL constraints.
The frontend supports modeling of DFDs through drag-and-drop, using three node types (I/O-, Storage-, and Function-Nodes), connected by directed flows from Input- to Output-Pins, as shown in the provided screencast\footnote{\url{https://youtu.be/L-PGdWoPtlw} (11.05.2026)}.
Label values from the data dictionary can be applied to nodes via drag-and-drop. The behavior of each Output-Pin can be defined using assignments in an in-place editor.
An exemplary excerpt of the frontend is shown in \autoref{fig:extended-dfd-example}.
Constraints are formulated using the DSL described in \autoref{sec:concepts}. Both the constraints DSL and assignment language feature syntax highlighting and autocompletion. Renaming labels or flows automatically updates all corresponding references.
In addition to modeling and analysis, the editor supports saving and loading DFDs in multiple formats, exporting diagrams as images or PDFs, and automatic layouting. It also provides readability features such as hiding flow and node labels in larger models.

The backend encapsulates the analysis functionality and is hosted on bwCloud, a federated Infrastructure-as-a-Service platform in Germany. 
Communication between the frontend and backend occurs via WebSocket\footnote{\url{https://datatracker.ietf.org/doc/html/rfc6455} (11.05.2026)}. When an analysis is triggered, the frontend packages the diagram data, along with the constraints and label information, into a JSON format and sends it to the backend. The backend processes the model, runs the analysis, and returns a JSON response that includes the modified diagram enriched with information about constraint violations and propagated labels, which are parsed and displayed to the user. %% Alex ~ 0.5
\section{Research Applications}
\label{sec:appl}
The \xdecaf framework has been used in several research approaches to achieve different goals, illustrating how it can be applied and customized to varying needs.

\textbf{External Validation:} 
To enable the integration of \acp{DFD} from existing research, \xdecaf has been extended to accept PlantUML files as input \cite{niehues2024integrating}.
The integration of externally authored \acp{DFD}, such as the \textsc{microSecEnD} dataset \cite{microSecEnD23}, serves as external validation for \xdecaf. The dataset is made up of \acp{DFD} derived from the source code of open-source microservice applications and manually created repair variants that fix information security concerns. Applying \xdecaf to all 132 modeled \ac{DFD} variants showed that only 115 variants exhibited expected results. A manual inspection of the 17 variants with diverging results revealed that none of the discrepancies stemmed from the transformation from PlantUML or the constraint formalization in \xdecaf, but rather could be traced back to faults in the manually created \textsc{microSecEnD} variants \cite{niehues2024integrating}. %These identified faults were subsequently reported to the authors of the dataset.
%the input of PlantUML files, as well as  input from two additional architecture design notations, \ac{PCM} and PlantUML models.
%\ac{PCM} \cite{palladioBook} is an \ac{ADL} for component-based software systems. 
%It is supported by a transformation that derives \acp{DFD} from a \ac{PCM} model and analyzes them for confidentiality violations \cite{seifermann2022detecting, huller2024towards}.
%Niehues et al.~\cite{niehues2024integrating} integrate externally authored \acp{DFD} expressed in PlantUML, as direct inputs to the analysis.
%Both extensions reuse the same analysis core, so any improvement to \xdecaf is immediately available across all supported input formats.

\textbf{Coupled and Downstream Analyses:}
Building on the support for \ac{PCM} instances, \citet{reiche2025detecting} use \xdecaf as a design-time analysis in a formalized composition approach of architectural and source code analyses. 
An analysis composition, using \xdecaf, was able to uncover an active encryption-related vulnerability in an open-source electric-vehicle charging-station architecture \cite{corallo2026everestdataset}.
%, providing field evidence that \xdecaf's design-time results are meaningful when combined with implementation-level findings.
\citet{boltz2024modeling} apply \xdecaf with \ac{PCM} instances to evaluate compliance to \ac{ZTA} principles, in particular access control and the principle of least privilege. As \acp{ZTA} can introduce bottlenecks in policy decision and enforcement, the approach combines Palladio design-time performance simulations \cite{palladioBook} and \xdecaf analysis to enable informed trade-off decisions between performance and adherence to \ac{ZTA} principles \cite{boltz2024modeling}. 
%Using the support for \ac{PCM} instances
%Combining the resulting confidentiality verdicts with the \ac{PCM} performance simulation~\cite{palladioBook} yields informed trade-off decisions between security and performance, an integration that is only possible because \xdecaf shares its underlying architectural model with the existing Palladio analysis stack.
% \medskip

% \noindent Beyond, three distinct lines of research are also based on \xdecaf:

\textbf{\acs{ABUNAI}:}
\ac{ABUNAI}\footnote{\url{https://github.com/abunai-dev/ABUNAI} (11.05.2026)} is a line of research focusing on design-time confidentiality analysis under uncertainty. Its central idea is to represent classified uncertainty as a first-class entity and explicitly handle it during analysis. It extends \xdecaf in two distinct approaches: First, an uncertainty impact analysis that extends the label propagation of \xdecaf and propagates uncertainty along the data flows to predict its impact on the system's confidentiality \cite{hahner2023impact}. Second, an uncertainty-aware confidentiality analyses that identify violations of confidentiality requirements with respect to uncertainty, by adding a preprocessing step, prior to label propagation and analysis, that creates variations of data flows that contain uncertainties \cite{hahner2023model}. How variations are created depends on the type of uncertainty and on the element it is associated with.

\textbf{\acs{arcovia}:}
\ac{arcovia}\footnote{\url{https://github.com/arcovia-dev/Mitigation} (11.05.2026)} closes the loop from identification of violations through analysis to mitigation of identified violations. 
Its first approach \cite{niehues2025architecture,niehues2025mitigation} builds on \ac{ABUNAI} and, using machine learning, ranks variation points resulting from uncertainty by how often they contribute to violations, then composes alternatives into a violation-free \ac{DFD}. 
Its second approach \cite{niehues2026efficient} encodes violations identified by \xdecaf and a set of admissible \ac{DFD} modifications as a SAT problem, which is solved to select a minimally invasive or user-preferred solution. 
Both approaches treat \xdecaf as their oracle for confidentiality. %, and their evaluations show that \xdecaf's results are consistent and machine-readable to drive automated mitigation options.

% \textbf{\acs{colja}:}
% The \ac{colja}\footnote{\url{https://github.com/colja-dev}} line of research focuses on enabling and improving compliance-related activities based on \acp{DFD}, to uses \xdecaf to bridge software architecture models and legal threat modeling~\cite{threatmodelingmanifesto}. 
% Boltz et al.~\cite{boltz2026continuous} place \xdecaf at the center of a model-driven workflow for ongoing interdisciplinary collaboration between legal and technical experts, where \xdecaf identifies system-level violations of formalized legal requirements. 
% Boltz et al.~\cite{boltz2025knowledge} complement this with a knowledge-transfer approach in which \xdecaf provides the technical anchor for legal concepts on architectural models. 
% MDPA thus validates that \xdecaf's constraint language is expressive enough to encode requirements outside its original confidentiality scope.

\textbf{\acs{colja}:}
The \ac{colja}\footnote{\url{https://github.com/colja-dev} (11.05.2026)} research line leverages and extends \xdecaf for compliance-driven threat modeling and interdisciplinary collaboration. 
Addressing the inherently interdisciplinary nature of legal compliance, \ac{colja} defines a model-driven workflow for interdisciplinary compliance checking, based on consistency-preserving, bidirectional transformations between legal viewpoints and \xdecaf \acp{DFD} \cite{boltz2026continuous}. Legal aspects are mapped to labels and analyzed with \xdecaf to identify potential compliance violations under change and legal uncertainty.
An additional approach supports legal knowledge transfer by attaching explanatory legal comments to \ac{DFD} elements, using \xdecaf constraints to identify their scope and extending the \xdecaf online editor to show the comments as tooltips \cite{boltz2025knowledge}.

%%Benny, Nils, Nicolas, Felix ~1
\section{Related Work}

%% introduce similar tools
%% position xDECAF

Most existing tooling for architecture-based security analysis focuses on a single security concern within a largely closed tool ecosystem. These approaches typically target specific tasks, such as threat modeling or confidentiality analysis, and offer limited extensibility beyond their predefined scope.
Within the domain of DFD-based approaches, the SecDFD tooling \cite{sion_solution-aware_2018, tuma_flaws_2019} extends traditional DFDs by incorporating security solutions to support structured threat elicitation. Similarly, industrial tools such as Microsoft’s Threat Modeling Tool \cite{shostack2008experiences} and OWASP Threat Dragon \cite{owasp-threat-dragon} rely on rule-based analyses with fixed threat libraries, limiting their applicability to a predefined set of threats.

Beyond DFD-based methods, other model-driven approaches such as UMLsec \cite{jurjens2002umlsec, peldszus_secure_2019} and CORAS \cite{lund2010model} employ UML-based architectural models to reason about security properties. While these approaches provide formalized mechanisms for analyzing specific concerns, they are typically tailored to particular analysis goals and modeling paradigms.
In contrast to these tools, \xdecaf decouples label types and labels from the analysis logic, enabling a broader range of analyses through its constraint-driven framework.
%In contrast to these tools, which address fixed analysis problems, \xdecaf enables a broader range of analyses through its constraint-driven framework. By decoupling the definition of label types and labels from the analysis logic, \xdecaf supports flexible, extensible data-flow analyses that can be adapted to diverse security concerns. \todo{der Absatz kürzen, zu viel Werbung für sich selbst}
%As demonstrated in Section~\ref{sec:appl}, this flexibility allows \xdecaf to cover a wider spectrum of analysis scenarios compared to existing approaches. %% Benny ~ 0.4
\section{Conclusion}
\label{sec:concl}

We presented \xdecaf, an open and extensible tool for architecture-based information-security analysis on \acp{DFD}.
\xdecaf couples an extended \ac{DFD} metamodel with a constraint \ac{DSL}, and a browser-based editor that makes the tool accessible for researchers as well as architects, security specialists, and developers.
Beyond the tooling, we also provided a curated catalog of over 20 \ac{DFD} models, each with documented constraints and expected violations, creating a baseline for future research.
%The tool is adopted in several research lines, ranging from microservices and \ac{PCM} models, through uncertainty-aware confidentiality analysis (ABUNAI) and automated mitigation (ARCoViA), to legal threat modeling (MDPA).

Current research uses \aclp{llm} to derive initial \acp{DFD} as well as \xdecaf labels and constraints from natural language requirements, enabling \xdecaf-based identification of \emph{potential} threats, and extends the \xdecaf editor with question-driven elicitation to guide experts in adding threat-relevant labels.
Additional efforts focus on integrating context models into the analysis and closing the loop by enabling \ac{arcovia} inside the web editor. %%Benny ~ 0.1

\begin{acks}
This work was supported by funding from the topic Engineering Secure Systems of the Helmholtz Association (HGF), KASTEL Security Research Labs, Karlsruhe, and the Deutsche Forschungsgemeinschaft (DFG, German Research Foundation) – CRC 1608 – 501798263.
We thank Manuel Córcoles and Elias Wörner for their contributions as research assistants.
\end{acks}

\bibliographystyle{ACM-Reference-Format}
\bibliography{sample-base}

%\appendix

\end{document}